\documentclass[]{spie}  

\usepackage{amsmath,amsfonts,amssymb}
\usepackage{graphicx}
\usepackage[colorlinks=true, allcolors=blue]{hyperref}
\usepackage{svg}
\usepackage{xcolor}
\usepackage{subcaption}
\usepackage[perpage]{footmisc} 
\usepackage{tabularx}

\usepackage{array}
\usepackage{cite}

\title{Pointing Model Meets Deep Learning: A Retrospective Study on a MeerKAT+ Telescope Applying Deep Learning Methods for Blind Pointing Corrections}

\author[a]{Stefan Thoms}
\author[a]{Matthias Reichert}
\affil[a]{OHB Digital Connect GmbH, Weberstrasse 21, 55130 Mainz, Germany}

\authorinfo{\hspace*{-0.2cm}Further author information: (Send correspondence to S.T.)\\ \hspace*{0.45cm}S.T.: E-mail: stefan.thoms@ohb.de, Telephone: +49-6131-2777-151\\ \hspace*{0.45cm}M.R.: E-mail: matthias.reichert@ohb.de, Telephone: +49-6131-2777-131}

\begin{document} 
\maketitle

\begin{abstract}
This study aims to compare the effectiveness of deep learning methods, specifically Feedforward Neural Networks (FNN), with traditional Pointing Models (PMs) for compensating Blind Pointing Errors in astronomical instruments. Ambitious projects like the ongoing study for the Atacama Large Aperture Submillimeter Telescope (AtLAST\footnote[2]{\href{https://www.atlast.uio.no/}{https://www.atlast.uio.no/}}) inspired to investigate possible improvements of traditional Pointing Error (PE) modeling. The study assesses the practicality of FNNs by applying them to data from an instrument in operation: a precursor MeerKAT+ telescope from the Max Planck Institute for Radio Astronomy (MPIfR) to extend the current MeerKAT Radio Telescope Array at the South African Radio Astronomy Observatory (SARAO) site in the Meerkat National Park in South Africa. 
\end{abstract}

\\
\section{INTRODUCTION}
\label{sec:intro} 
For telescopes, Pointing Errors (PE) are a critical factor that affects the accuracy and efficiency of signal reception. This is particularly pertinent to radio telescopes, which commonly lack feedback instrumentation for real-time pointing estimation and closed-loop PE compensation with celestial reference sources. PEs originate from a variety of sources, including thermal inhomogeneities in the telescope's structure, other environmental factors such as wind and humidity, and distortions caused by gravitational effects. Additionally, mechanical misalignments and the inherent limitations in mechanical manufacturing or control loops also contribute to PEs, which vary depending on the telescope's pointing direction and the environmental conditions.

Typically, models for predicting these PEs primarily rely on standard Pointing Models (PMs), which utilize numerous geometric and physical correction factors.\cite{Greve1996, guiar1987antenna, keitzer1991deterministic, meeks2004sources, de2007deconstructing, white2022green} However, advanced computational machine-learning methods, such as FNNs, are beginning to offer new prospects for enhancing the precision of PE predictions.\cite{nyheim2024machine, chen2022improved} This study aims to evaluate the capability and effectiveness of FNNs compared to a basic conventional PM in a proof-of-principle setup. The primary motivation is to explore the potential of FNNs in practice, assessing their strengths and limitations when trained on a limited dataset from a telescope in operation. The dataset used for this study originates from a prototype called \textsc{SKA MPI Demonstrator} from the Max Planck Institute for Radio Astronomy (MPIfR), with OHB~DC as the main contractor. It was built at the South African Radio Astronomy Observatory (SARAO) site in the MeerKAT National Park to gain experience with the previous design of the dish structure, serving as a precursor for 14 additional telescopes to extend the MeerKAT Radio Telescope Array at the SARAO. Integration into the SKA-MID (Square Kilometre Array - MID) observatory is planned at a later stage as the telescope design follows the current SKA-MID requirements.

An additional motivation was to prioritize ease of implementation by utilizing open-source libraries and ensuring that the computational requirements remain within the capabilities of standard computers, rather than relying on powerful supercomputing resources.
\\
\section{Pointing Error Modeling}
The forthcoming subsections will describe the two considered modeling approaches: the theoretically based Pointing Model (PM) and the machine learning framework, a Feedforward Neural Network (FNN). Each method was evaluated using the dataset from the MeerKAT+ telescope (see Section~\ref{ssec:data_background}) and was fully conducted in \textsc{Python}.
\\
\subsection{Pointing Model (PM)}
\label{ssec:pm}
The PM used within this study is well-established in various variations and is derived from a publication by Greve et. al for the IRAM 30M telescope.\cite{Greve1996, guiar1987antenna, keitzer1991deterministic, meeks2004sources, de2007deconstructing, white2022green}
\begin{align}
\Delta AZ(AZ,EL) &=  C_{\text{AZ}} + C_{\text{XEL}}\sec{\left(EL\right)} + P_{\perp}\tan{\left(EL\right)} + T_{\text{EW}}\cos{\left(AZ\right)}\tan{\left(EL\right)} + T_{\text{SN}}\sin{\left(AZ\right)}\tan{\left(EL\right)} \label{equ:stdPM_dAZ} \\
\Delta EL(AZ,EL) &= C_{\text{EL}} - T_{\text{EW}}\sin{\left(AZ\right)} + T_{\text{SN}}\sin{\left(AZ\right)} + TS^{1}_{\text{EL}}\sin{\left(EL\right)} + TC^{1}_{\text{EL}}\cos{\left(EL\right)}\label{equ:stdPM_dEL}
\end{align}
This model is tailored for any two-axis telescope with an azimuth and an elevation axis. Standard PMs are typically (partially) based on analytical terms to compensate for geometric systematic errors caused by misalignments and finite manufacturing precision. Parameters C$_i$ compensate for static offsets ($i$=AZ: Azimuth, $i$=EL: Elevation, $i$=XEL: Cross-Elevation), P$_{\perp}$ for the lack of orthogonality between azimuth and elevation axes, and T$_j$ for misalignments of the azimuth axis ($j$=EW: East-West, $j$=NS: North-South). Additionally, generic harmonic terms (here TS$^{1}_{\text{EL}}$/TC$^{1}_{\text{EL}}$) are often introduced, to correct for effects of a more complex nature, such as effects of bearing runouts or elevation position-dependent gravitational sag of the telescope. 
\\
\subsection{Feedforward Neural Networks (FNN)}
\begin{table}[tb]  
\centering  
\caption{List of preselected, invariable hyperparameters of Scikit-learn's solver \textsc{MLPRegressor} used for training all FNNs. Only hyperparameters relevant to the chosen solver \textit{adam} are listed. For further information, refer to Scikit-learn's documentation \cite{sklearnMLPRegressor}.}
\label{tab:invar_hyperparameter}
\begin{tabularx}{\textwidth}{p{3cm}p{3cm}X}
\hline
\hline
Hyperparameter & \centering Argument & Comment\\
\hline
activation                  & \centering relu        & Rectified Linear Unit \\
solver                      & \centering adam        & Stochastic Gradient-Descent Optimiser\cite{kingma2014adam}\\
alpha                       & \centering 10$^{-3}$   & Regularization term\\
batch$\_$size               & \centering auto        & Minibatches for stochastic optimizers\\
learning$\_$rate$\_$init    & \centering 0.05        & Initial learning rate  \\
max$\_$iter                 & \centering 10$^4$      & For solver \textit{adam}: Number of epochs \\
shuffle                     & \centering True        & Shuffling of samples in each iteration\\
random$\_$state             & \centering None        & None: Random initial values for weights and biases\\
tol                         & \centering 10$^{-4}$   & Tolerance for optimization~/ convergence evaluation \\
validation$\_$fraction      & \centering 0.1         & The proportion of training data to set aside as validation set for early stopping\\
early$\_$stopping           & \centering True        & True: Termination of training when validation score is not improving any more \\
beta$\_$1                   & \centering 0.9         & Exponential decay rate for estimates of first moment vector \\
beta$\_$2                   & \centering 0.999       & Exponential decay rate for estimates of second moment vector\\
epsilon                     & \centering 10$^{-8}$   & Value for numerical stability \\
n$\_$iter$\_$no$\_$change   & \centering 30          & Maximum number of epochs to not meet tol improvement\\
\hline
\hline
\end{tabularx}
\end{table}
In recent years, the availability of several open-source libraries has greatly facilitated the implementation of machine learning approaches. Among the most widely used libraries are \textsc{TensorFlow}, \textsc{PyTorch} and \textsc{Scikit-learn}. In this study, we utilized \textsc{Scikit-learn} version 0.24.1.\\
These libraries offer numerous implemented solvers with various hyperparameters to adjust. Ideally, comprehensive testing of various combinations of hyperparameters within a wide range would be performed to identify the best-performing configuration.
However, this quickly demands significant computational time and power, exceeding standard computational capabilities for practical purposes. Therefore, we performed a preliminary selection of some hyperparameters through testing with several samples to minimize the number of hyperparameter combinations and, hence, the computational time required. Table~\ref{tab:invar_hyperparameter} lists the selected standard grid of hyperparameters along with their default values. Whereas Table~\ref{tab:var_hyperparameter} provides the set of variable hyperparameters, specifically the layer sizes and the number of deep layers tested. \\
For the optimization of the parameters within the FNN (weights and biases), we used the built-in \textsc{MLPRegressor} (\textsc{Multi-Layer Perceptron Regressor}) from \textsc{Scikit-learn}. The input layer had no activation function and the parameters were optimized to minimize the squared error (half of the mean squared error). The standard set of invariable hyperparameters used for all trained FNNs is listed in Table~\ref{tab:invar_hyperparameter}.
\\
\begin{table}[tb]  
\centering  
\caption{This table lists the selected hyperparameter, \textit{hidden$\_$layer$\_$sizes}, used by the \textsc{Scikit-learn} \textsc{MLPRegressor} for training the FNNs. It details all configurations of \textit{hidden$\_$layer$\_$sizes}, resulting in 43 distinct architectures for the FNNs to evaluate. For additional details, refer to the \textsc{Scikit-learn} documentation \cite{sklearnMLPRegressor}.}
\label{tab:var_hyperparameter}
\begin{tabularx}{1\textwidth}{Xp{4cm}p{4cm}p{4cm}}
\hline
\hline
Hyperparameter            & \centering Number Hidden Layer & \centering Size Range &  Number of Sizes  \\
\hline
hidden$\_$layer$\_$sizes  & \centering 1                   &  \centering [10, 100] &  \hspace{1cm} 20              \\
hidden$\_$layer$\_$sizes  & \centering 2                   &  \centering [10, 100] &           \hspace{1cm} 20              \\
hidden$\_$layer$\_$sizes  & \centering 3                   &  \centering [5, 20]   &          \hspace{1.1cm}  3               \\
\hline
\hline
\end{tabularx}
\end{table}

\section{Dataset}
\subsection{Data Background}
\label{ssec:data_background}
\begin{figure}[b] 
  \centering 
  \includegraphics[width=1.0\linewidth]{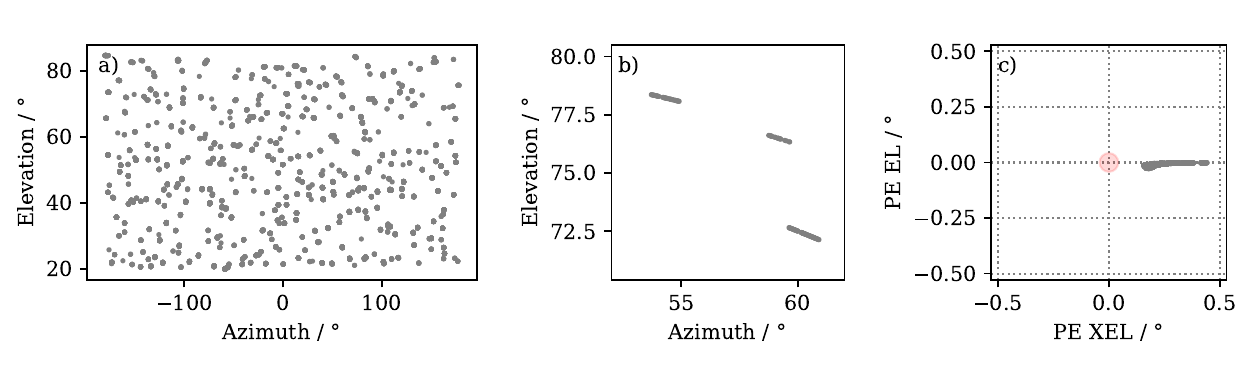} 
  \caption{a): Overview of all 409 tracks used as input for this study, illustrating sufficient coverage of the range of the axes. b): Enlarged detail of Figure a). c): Plot of all PEs prior to compensation using one of the proposed PE models. PE$^{\text{XEL}}_{\text{rms}}$~= 743~arcsec, PE$^{\text{EL}}_{\text{rms}}$~= 56~arcsec.}
  \label{fig:data_selection_ska} 
\end{figure}
To effectively compensate for (quasi-) static Pointing Errors, precise estimation is imperative, ideally covering the entire operational range of the telescope’s axes. It is essential to emphasize that only comprehensive reference data enables the identification of detailed systematics and correlations between the targeted and actual observed pointing positions, and subsequently compensate for them. This process is indispensable for any form of modeling.

For the \textsc{MeerKAT+} telescope, PEs could be accurately estimated using bright reference stars from a star catalogue and surveyed by an optical camera \cite{glaubach2024verifying}. This camera has been mounted and calibrated above the elevation axis of the telescope near the center of the main dish. Being mounted in the main dish introduces a residual discrepancy between the pointing of the camera and the pointing of the telescope's radio receivers. However, this study methodically investigates the modeling of PEs, and this discrepancy does not undermine the resulting aspects of the analysis. The evaluation involves assessing the PE in deviations of the elevation position ($\Delta$EL) and cross-elevation positions ($\Delta$XEL), the latter is also expressible as a deviation of the azimuth position ($\Delta$AZ~= $\Delta$XEL$\cdot$sec(EL)). 

It should be noted, that a high-accuracy tiltmeter with a repeatability of 0.1 arcsec is installed below the elevation axis. As a result, it rotates with the azimuth position and is not affected by changes in the elevation position. The data from the tiltmeter is used to compensate for tilts of the azimuth axis in real-time, based on physical models for the position of the tiltmeter. In this study, the modeling of the data shown in the results is performed post-tiltmeter compensation, thus, the modeled PEs already include compensations based on the tiltmeter data.

The complete dataset is based on a survey campaign conducted in September 2021 and was recorded over seven nights within three weeks to evaluate the pointing model of the \textsc{MeerKAT+} telescope. The pre-filtered dataset comprises 746 tracks, each approximately $\approx$~100~s in length.
\\
\subsection{PRE-PROCESSING}
\label{sec:data_preprocessing}
\subsubsection*{Data Reduction~/ Filtering}
A subsidiary objective of the study was to minimize the effort expended on data preparation, aiming to utilize the input data mainly in its unadulterated form. This approach was intended to yield a more generalized understanding of performance in real applications and with in-the-field experimental data. The focus was deliberately shifted away from a rigorously curated subset of data most amenable to modeling, to better reflect the challenges and dynamics encountered in actual operational environments. 

Nonetheless, some fundamental processing and preparations, i.e., filtering, of the input data were necessary to achieve both robust and optimized results. Therefore, to exclude mainly data with dominant dynamic PE fraction, the data was filtered for wind speeds not exceeding 5 m/s and further for elevation angles between 10° and 85°. To reduce the necessary computational performance, the original sampling rate of 10~Hz was reduced to 0.5~Hz. Accordingly, averaging was performed over a two-second window. Filtering of the dataset resulted in a final set of 409 tracks, comprising an average of approximately 24 data points per track. The remaining tracks are shown in Figure~\ref{fig:data_selection_ska} together with the related PEs of each data point before a PM or FNN was applied. 
\\
\subsubsection*{Train-Test Splitting}
\begin{figure}[t] 
  \centering 
  \includegraphics[width=0.99\linewidth]{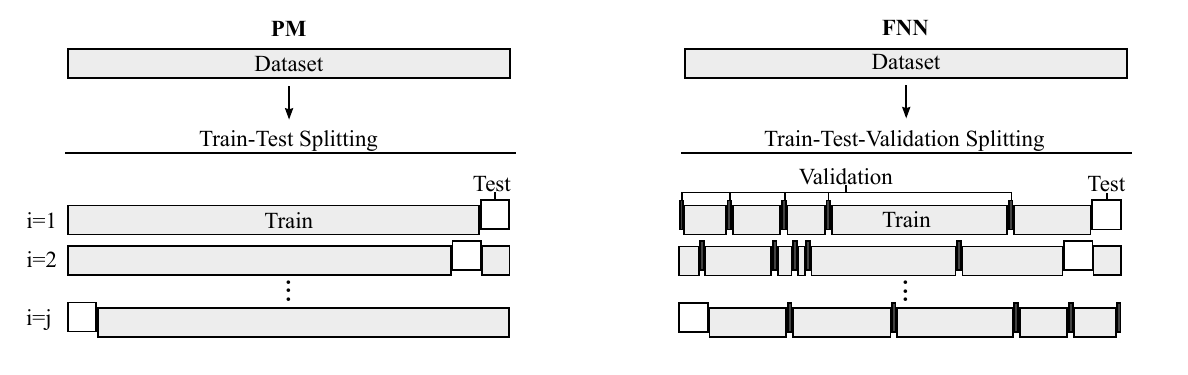} 
  \caption{Comparison of the PM and FNN model evaluation processes. The PM model was evaluated using a straightforward Train-Test Split approach, with several iterations of training followed by testing. In contrast, the FNN model employs a Train-Test-Validation Split, which splits an additional validation fraction to first optimize the model's parameters (weights and biases) and then test the model for each hyperparameter (see Table~\ref{tab:var_hyperparameter}) on the test fraction.} 
  \label{fig:train_test_splitting2} 
\end{figure}
Dealing with relatively limited dataset sizes, especially for training an FNN, small fractions for testing are preferred, to provide as much data for training as possible while simultaneously achieving robust and comprehensive testing on a statistical basis. Consequently, leave-one-out cross-validation (LOOCV) was considered. However, LOOCV incurs high computational costs and is time-consuming, respectively. As a compromise, testing fractions of 2.5~\% were utilized to evaluate both the PM and the FNN across 40 cross-validation folds. Figure~\ref{fig:train_test_splitting2} illustrates the Train-Test Splitting procedure for evaluating the PM along with the hyperparameters and parameters of the FNN. For the FNNs, fractions for validation (10~$\%$) were randomly split from the train set to evaluate the parameter on one hyperparameter configuration (e.g. iteration $i$=1). This was then done for each of the 43 hyperparameter configurations (see Table~\ref{tab:var_hyperparameter}) on each Train-Test Splitting fold (each $i$). To evaluate the best-performing hyperparameter configuration, the mean best-performing configuration over all Test-Train Splittings was identified. 
\\
\subsubsection*{Transformation}
For modeling the data with the PM, no scaling of the input data (azimuth and elevation position) was applied, as most of the terms of the model consist of trigonometric functions and, hence, are based on the angular framework.  
For training the FNN, each feature of input data $x$ is scaled by Z-Score Normalization with $x_{sc} = (x-\mu)/\sigma$, with $\mu$ the mean and $\sigma$ the standard deviation.
\\
\subsubsection*{Feature Selection}
FNNs offer the advantage that feature selection can generally be performed by evaluating the FNN's parameters (weights and biases), hence, by the FNN itself. However, an efficient feature selection can be beneficial, especially to optimize for computational time and limited data to train on. While certain features, such as data from temperature sensors, are predestined to be a high-impact source of PEs, the original data for each track contains many more features to select from, including not only multiple temperature sensors but also further environmental data,  axis velocities, motor torques, and motor positions.
\begin{table}[t]
    \caption{The table displays potential features as input for FNNs, ranked by Mutual Information (MI) (top) and Pearson Correlation Coefficient (PCC) (bottom) scores, evaluated against PEs (PEs). MI is tested against the RSS PE of $\Delta$EL and $\Delta$XEL, whereas PCC is tested against $\Delta$EL and $\Delta$XEL, separately. Values are organized in descending order based on their score of the shown PE parameter to identify the most potent inputs for training an FNN to enhance predictive accuracy regarding PEs. v$_i$: velocity of $i$'s axis, RH$_\text{mws}$: relative humidity of near-by metrology weather station, T$_j$: temperature sensor data of $j$'s location.}
    \label{tab:ranked_MI_PCC}
     \vspace{0.1cm}
    \begin{tabularx}{\textwidth}{p{2cm}|p{3.2cm}p{3.2cm}|p{3.2cm}p{3.2cm}X}
        &\multicolumn{2}{c}{Mutual~Information}& \multicolumn{2}{|c}{Pearson Correlation Coefficient}&\\
        \hline
        \hline
        \centering Ranking    & \centering Feature            & \centering PE Parameter &  \centering Feature            & \centering PE Parameter& \\ 
        \hline
        \centering 1      & \centering v$_{\text{AZ}}$   & \centering PE RSS   & \centering Torque$_{\text{EL}}$    & \centering PE XEL & \\
        \centering 2      & \centering RH$_\text{mws}$          & \centering PE RSS   & \centering RH$_\text{mws}$         & \centering PE EL &\\
        \centering 3      & \centering T$_{\text{mws}}$  & \centering PE RSS   & \centering v$_{\text{EL}}$  & \centering PE EL &\\
        \centering 4      & \centering v$_{\text{EL}}$   & \centering PE RSS   & \centering T$_{\text{1}}$   & \centering PE EL &\\
        \hline
        \hline
    \end{tabularx}
\end{table}
To find a simple and fast systematic approach, sets of features were evaluated for both Mutual Information (MI) \cite{shannon1948mathematical}$^,$\cite{kreer1957question} and for the Pearson Correlation Coefficient (PCC) \cite{pearson1896vii} with respect to Pointing Errors, aiming to uncover dependencies prior to training an FNN. This approach originates from the idea of identifying and prioritizing features that contain correlations, thereby providing beneficial input for a Pointing Model to predict Pointing Errors. Table~\ref{tab:ranked_MI_PCC} lists the top four features resulting from both methods of dependency investigation. The set positions for both axes (Set$_{\text{AZ}}$, Set$_{\text{EL}}$) are excluded, as they are known to be highly ranked and are consistently used as input for every model. The derived Feature Sets (FS) are documented in Table~\ref{tab:feature_cases}. Additionally, an FS comprising all potential and available input features (58 inputs) and a set incorporating all available temperature data (added to position of azimuth Set$_{\text{AZ}}$ and elevation Set$_{\text{EL}}$) are listed and were tested.
\begin{table}[t]  
\centering  
\caption{Various specified Feature Sets (inputs) will be used to generate predictions (outputs) with PM and FNNs. Set$_{\text{AZ}}$ and Set$_{\text{EL}}$ represent the set position for the azimuth and elevation axes within the control loop of the telescope. v$_{axis}$ denotes the axes' velocities, RH$_{\text{mws}}$ and T$_{\text{mws}}$ are measures of relative humidity and temperature, respectively, provided by a nearby meteorology weather station, whereas T$_{\text{j}}$ stands for temperature sensors mounted at the azimuth pedestal and T$_{\text{Tilt}}$ for a sensor within the installed tiltmeter above the azimuth pedestal. Additionally, Torque$_\text{EL}$ is the applied torque of one of the motors for the elevation axis.}
\label{tab:feature_cases}
\begin{tabularx}{1\textwidth}{p{3.5cm}p{8.5cm}>{\centering\arraybackslash}p{4cm}X}
\hline
\hline
Feature Set                   &   Features~/ Input                                  &  Prediction~/ Output     &                \\
\hline
FS~P     &   Set$_{\text{AZ}}$, Set$_{\text{EL}}$                                                               & $\Delta$AZ, $\Delta$EL &  \\
\hline
MI~1     &   Set$_{\text{AZ}}$, Set$_{\text{EL}}$, v$_{\text{AZ}}$                                              & $\Delta$AZ, $\Delta$EL & \\
MI~2     &   Set$_{\text{AZ}}$, Set$_{\text{EL}}$, v$_{\text{AZ}}$, RH$_{\text{mws}}$                                    & $\Delta$AZ, $\Delta$EL & \\
MI~3     &   Set$_{\text{AZ}}$, Set$_{\text{EL}}$, v$_{\text{AZ}}$, RH$_{\text{mws}}$, T$_{\text{mws}}$                  & $\Delta$AZ, $\Delta$EL & \\
MI~4     &   Set$_{\text{AZ}}$, Set$_{\text{EL}}$, v$_{\text{AZ}}$, RH$_{\text{mws}}$, T$_{\text{mws}}$, v$_{\text{EL}}$ & $\Delta$AZ, $\Delta$EL & \\
\hline
PC~1      &  Set$_{\text{AZ}}$, Set$_{\text{EL}}$, Torque$_{\text{EL}}$                                                & $\Delta$AZ, $\Delta$EL  &\\
PC~2      &  Set$_{\text{AZ}}$, Set$_{\text{EL}}$, Torque$_{\text{EL}}$, RH$_{\text{mws}}$                                      & $\Delta$AZ, $\Delta$EL  &\\
PC~3      &  Set$_{\text{AZ}}$, Set$_{\text{EL}}$, Torque$_{\text{EL}}$, RH$_{\text{mws}}$, v$_{\text{EL}}$                     & $\Delta$AZ, $\Delta$EL  &\\
PC~4      &  Set$_{\text{AZ}}$, Set$_{\text{EL}}$, Torque$_{\text{EL}}$, RH$_{\text{mws}}$, v$_{\text{EL}}$, T$_{\text{1}}$     & $\Delta$AZ, $\Delta$EL  &\\
\hline
FS T & Set$_{\text{AZ}}$, Set$_{\text{EL}}$, T$_{\text{1}}$, T$_{\text{2}}$, T$_{\text{3}}$, T$_{\text{mws}}$, T$_{\text{Tilt}}$ & $\Delta$AZ, $\Delta$EL  &\\
FS X      &  Set$_{\text{AZ}}$, Set$_{\text{EL}}$, \textit{Set of 56 Parameters}     & $\Delta$AZ, $\Delta$EL  &\\
\hline
\hline
\end{tabularx}
\end{table}
\\
\section{RESULTS}
\label{sec:results}
The results are divided into two subsections. The first section compares the performance of the introduced Pointing Model (PM) (see Section~\ref{ssec:pm}) with a Feedforward Neural Networks (FNN), trained using various Feature Sets (FS) as input (see Section~\ref{sec:data_preprocessing} \textsc{Feature Selection}). The basic PM serves as a benchmark value, although it consists of only a few geometric terms and two input variables (Set$_{\text{AZ}}$, Set$_{\text{EL}}$). Consequently, while performance comparisons may not be equitable for FNNs with multiple input features, the PM shall be considered as a basic performance benchmark. The second part examines the performance of the PM and the FNN when the data used for the evaluation of the PM and the training of the FNN is reduced.
\\
\subsection{Feature Sets - Model Performances}
\subsubsection{Model complete PE with an FNN}
\begin{figure}[ptb] 
  \centering 
  \includegraphics[width=1.0\linewidth]{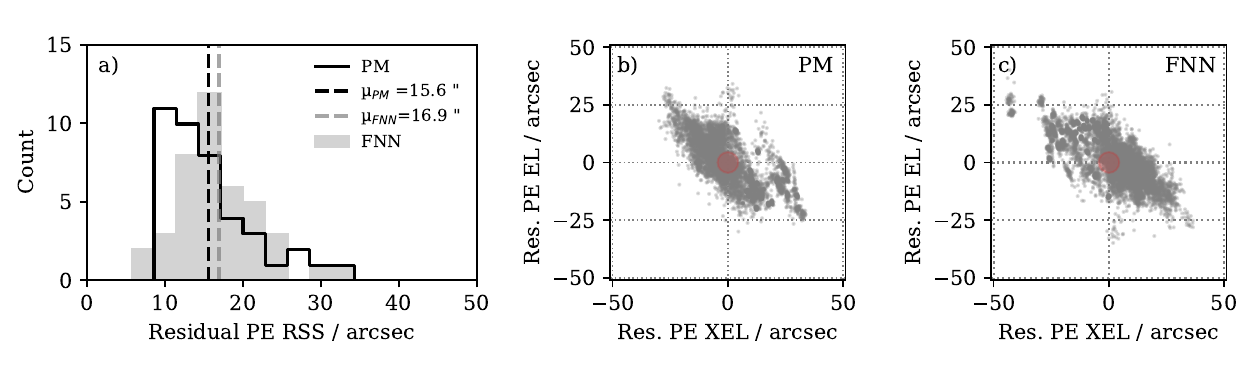} 
  \caption{All figures relate to results for FS~P (see Table~\ref{tab:feature_cases}). a): Resulting residual PE performances, after modeling initial PEs (see Figure \ref{fig:data_selection_ska} c)) with PM and FNN. Histograms show the root-mean-squared RSS values of PE in cross-elevation and elevation. Each count in each histogram represents the result of one of 40 test-train folds, as a result of the chosen 2.5~\% test splitting (see Section~\ref{sec:data_preprocessing}). The related mean values of the distributions $\mu_{model}$ are shown (see also Table~\ref{tab:all_mean_results}) b): Exemplary residual PEs of the complete dataset after modeling with PM. c): Residuals for modeling with the best-performing FNN of FS~P applied to the complete dataset. For comparison: Residuals prior to modeling, see Figure~\ref{fig:data_selection_ska} c).}
  \label{fig:result_ska_F11} 
\end{figure}

According to the FS - listed in Table~\ref{tab:feature_cases} - FNNs were tested in terms of their capabilities to model the complete PE, i.e., to potentially substitute the PM. First, for an equitable comparison, PM and FNN were tested with FS~P, modeling the PE solely based on the axes' positions Set$_{\text{AZ}}$ and Set$_{\text{EL}}$ as input features. Figure~\ref{fig:result_ska_F11}~a) shows a histogram of the resulting PEs, post-compensation with the evaluated and trained PM or FNN, respectively, of all Test-Train folds. Figure~\ref{fig:result_ska_F11}~b) and c) display the residual PEs of all surveyed data points for each model (PM~/ FNN) after the related PE compensation, in contrast to the PEs prior to compensation shown in Figure~\ref{fig:data_selection_ska}~c). 

\begin{figure}[pb] 
  \centering 
  \includegraphics[width=1.0\linewidth]{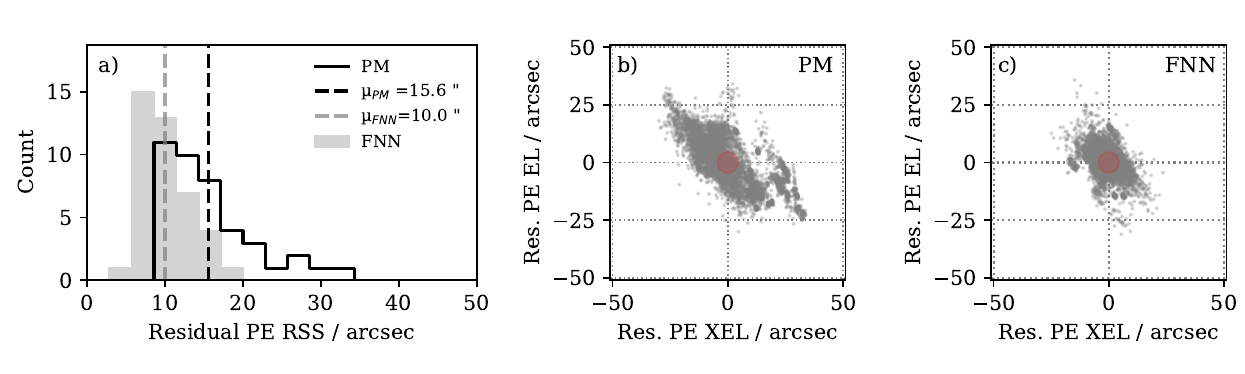} 
  \caption{Figures analogue to Figure~\ref{fig:result_ska_F11}. a): Resulting residual PE performances, after modeling initial PEs (see Figure \ref{fig:data_selection_ska} c)) with PM and FNN. PM is based on FS~P and FNN trained on PC~4, respectively. Histograms show the root-mean-squared RSS values of PE in cross-elevation and elevation. Each count in each histogram represents the result of one of 40 test-train folds, as a result of chosen the 2.5~\% test splitting (see Section~\ref{sec:data_preprocessing}). The related mean values of the distributions $\mu_{model}$ are shown (see also Table~\ref{tab:all_mean_results}) b): Exemplary residual PEs of the complete dataset after modeling with PM. c): Residuals for modeling with the best-performing FNN of FS~P applied to the complete dataset. For comparison: Residuals prior to modeling, see Figure~\ref{fig:data_selection_ska} c).}
  \label{fig:result_ska_FS9} 
\end{figure}

All numerical results of FS-related performances are provided in Figure~\ref{fig:result_all_FS} and Table~\ref{tab:all_mean_results} (column \textit{PM~/ FNN}). For an exemplary comparison using a more comprehensive FS as input for the FNN, Figure~\ref{fig:result_ska_FS9} shows identical resulting performance for the PM, but with the FNN utilizing an extended input FS: PC~4. The increased number of inputs for the FNN convincingly illustrates the improved performance in both the histogram (Figure~\ref{fig:result_ska_F11} a)) and the residual PEs (Figure~\ref{fig:result_ska_F11} c)).
\\
\subsubsection{Model Residual PE of PM with FNN}
\begin{figure}[ptb] 
  \centering 
  \includegraphics[width=1.0\linewidth]{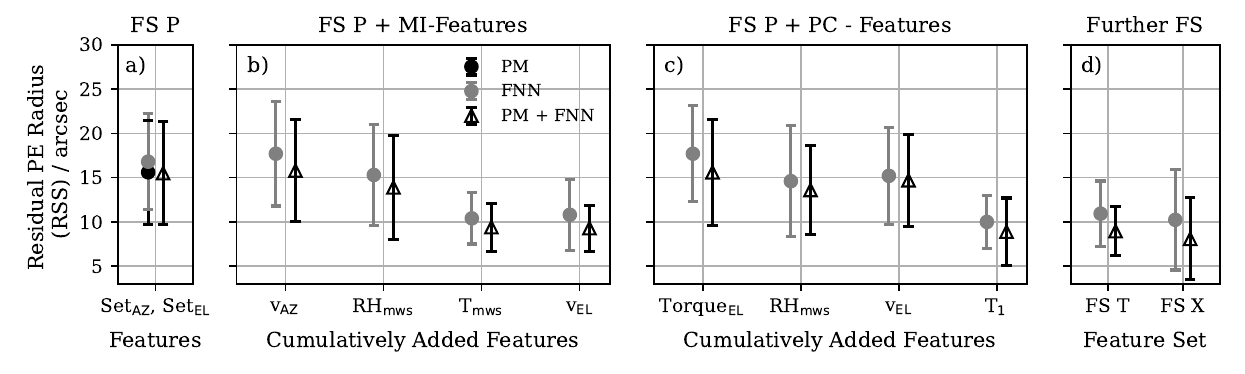}
  \caption{Visual illustration of resulting residual PE performances after compensations with PM and FNNs as well as for the joint performance of PM+FNN. Plots depict the results of various Feature Sets and reflect all numerical results from Table~\ref{tab:all_mean_results}. a): Results of PM, FNN and PM+FNN for FS~P. b): In addition to FS~P, features are cumulatively added in the order of top-ranked Mutual Information features (see Table~\ref{tab:ranked_MI_PCC}). c): In addition to FS~P, features are cumulatively added in the order of top-ranked Pearson Correlation features (see Table~\ref{tab:ranked_MI_PCC}). d): Further, FSes (see Table~\ref{tab:feature_cases}). Legend from b) applies to all subplots.}
  \label{fig:result_all_FS} 
\end{figure}

Given that the PM encapsulates a substantial amount of valuable information about the initial PEs, it seems logical to strive for a merger of the PM and FNN approaches. Hence, in this study not only it was tested to model the complete PEs with an FNN, but also the residual PEs after compensation with the PM (FS~P). Hence, first, the complete PE is modeled with the PM based on FS~P. The residual PEs (Figure~\ref{fig:result_ska_F11} b)) are then used as the output of the FNN to train on. The results of all tested FS for the FNNs are plotted in Figure~\ref{fig:result_all_FS} and the numerical final performances are listed in Table~\ref{tab:all_mean_results} (column \textit{PM~+ FNN}).
\begin{table}[pb]  
\centering  
\caption{Residual PEs after compensating for the initial PEs using a PM, an FNN, or a combined PM+FNN approach are presented. The column \textit{Features} lists the inputs utilized for evaluating and training the respective \textit{Method}. The column \textit{PM~/ FNN} provides residual PEs when the PM or FNN models the complete initial PE. The column \textit{PM~+ FNN} specifies the resulting performance for a joint approach, i.e., PEs are first modeled with PM (FS~P), and then an FNN is trained on the residual PEs of the PM for various FSes. The last column provides the difference between the listed performances of both methods.}
\label{tab:all_mean_results}
\begin{tabularx}{1\textwidth}{m{1cm}m{6.6cm}|Xm{2.41cm}|m{2.41cm}|m{1.0cm}}
&& \multicolumn{2}{c|}{PM~/ FNN}& \centering PM~+ FNN &\\
\hline
\hline
FS &  Features                                                                                            &  \centering Method   &  \centering  {Residual PE (RSS)~/ arcsec}  &   Residual PE (RSS)~/ arcsec   &  $\Delta$~/ arcsec \\
\hline 
FS~P  &  Set$_{\text{AZ}}$, Set$_{\text{EL}}$                                                                 & \centering   PM &\centering {15.6~$\pm$~5.9}     & \centering -  &\\
FS~P  &  Set$_{\text{AZ}}$, Set$_{\text{EL}}$                                                                 & \centering  FNN &\centering {16.9~$\pm$~5.4}     & \centering15.5~$\pm$~5.8 & (-1.4)\\
\hline
MI~1  &  Set$_{\text{AZ}}$, Set$_{\text{EL}}$, v$_{\text{AZ}}$                                                &  \centering FNN &\centering {17.7~$\pm$~5.9}     & \centering15.8~$\pm$~5.8   & (-1.9)\\
MI~2  &  Set$_{\text{AZ}}$, Set$_{\text{EL}}$, v$_{\text{AZ}}$, RH$_{\text{mws}}$                                      &  \centering FNN &\centering {15.3~$\pm$~5.7}     & \centering13.9~$\pm$~5.9   & (-1.4)\\
MI~3  &  Set$_{\text{AZ}}$, Set$_{\text{EL}}$, v$_{\text{AZ}}$, RH$_{\text{mws}}$, T$_{\text{mws}}$                    & \centering  FNN &\centering {10.4~$\pm$~2.9}     & \centering \hspace{0.17cm}9.4~$\pm$~2.7   & (-1.0)\\
MI~4  &  Set$_{\text{AZ}}$, Set$_{\text{EL}}$, v$_{\text{AZ}}$, RH$_{\text{mws}}$, T$_{\text{mws}}$, v$_{\text{EL}}$   & \centering  FNN &\centering {10.8~$\pm$~4.0}     & \centering \hspace{0.17cm}9.3~$\pm$~2.6  &(-1.5)\\
\hline
PC~1   &  Set$_{\text{AZ}}$, Set$_{\text{EL}}$, Torque$_{\text{EL}}$                                           &  \centering FNN &\centering {17.7~$\pm$~5.4}     & \centering 15.5~$\pm$~6.0   &  (-2.2)\\
PC~2  &  Set$_{\text{AZ}}$, Set$_{\text{EL}}$, Torque$_{\text{EL}}$, RH$_{\text{mws}}$                                 &  \centering FNN &\centering {14.6~$\pm$~6.3}     & \centering 13.6~$\pm$~5.0   &   (-1.0)\\
PC~3  &  Set$_{\text{AZ}}$, Set$_{\text{EL}}$, Torque$_{\text{EL}}$, RH$_{\text{mws}}$, v$_{\text{EL}}$                &  \centering FNN &\centering {15.2~$\pm$~5.5}     & \centering 14.1~$\pm$~6.0   &   (-1.1)\\
PC~4  &  Set$_{\text{AZ}}$, Set$_{\text{EL}}$, Torque$_{\text{EL}}$, RH$_{\text{mws}}$, v$_{\text{EL}}$, T$_{\text{1}}$&  \centering FNN &\centering {10.0~$\pm$~3.0}     & \centering\hspace{0.17cm}8.9~$\pm$~3.8  &   (-1.1)\\
\hline
FS T & Set$_{\text{AZ}}$, Set$_{\text{EL}}$, T$_{\text{1}}$, T$_{\text{2}}$, T$_{\text{3}}$, T$_{\text{mws}}$, T$_{\text{Tilt}}$ & \centering FNN & \centering 10.9~$\pm$~3.7 & \centering \hspace{0.17cm}9.0~$\pm$~2.8 & (-1.9) \\
FS X &  Set$_{\text{AZ}}$, Set$_{\text{EL}}$, \textit{Set of 56 Parameters}  & \centering FNN &\centering 10.2~$\pm$~5.7 &\centering \hspace{0.17cm}8.1~$\pm$~4.6 & (-2.1)\\
\hline
\hline
\end{tabularx}
\end{table}

\subsection{Performance on Data Reduction}
\label{ssec:data_reduction}
\begin{figure}[tb] 
  \centering 
  \includegraphics[width=1.0\linewidth]{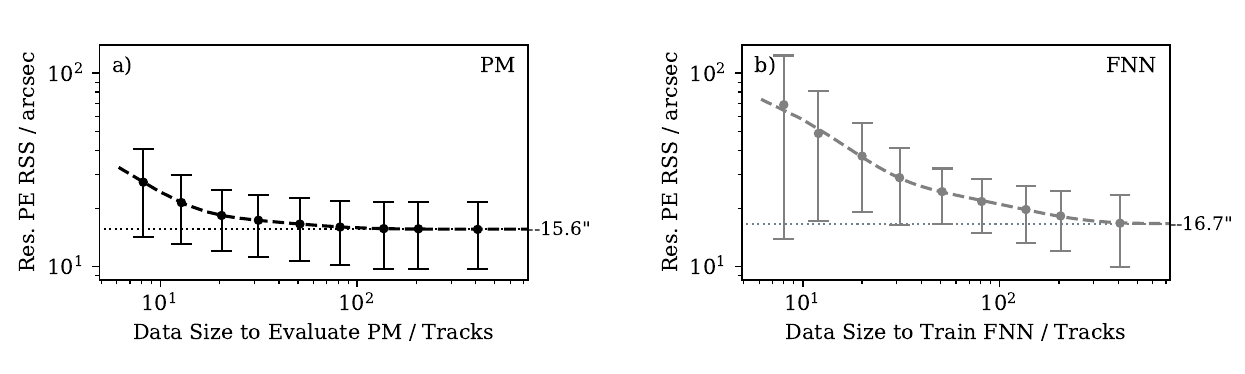} 
  \caption{Comparison of residual PEs based on the RSS of $\Delta$XEL and $\Delta$EL between the performance of the PM (a) and an FNN (b) as a function of data size for the evaluation and training, respectively. The values in both plots were fitted with a phenomenological double-exponential decay ($f(x)=a+b\exp(-cx)+d\exp(-ex)$), serving as a guide to the eye for potential performance convergence.}
  \label{fig:result_scarcity} 
\end{figure}

When applying any model to a real telescope, it is crucial to consider the amount of data, i.e., observing time, necessary for the evaluation/training of the model to reach a specific performance. Therefore, an exemplary comparison of both models (PM/FNN) predicting the complete PE on FS~P (only AZ and EL as input) was conducted using artificially reduced data for training each model. Figure~\ref{fig:result_scarcity} shows the resulting performance of the PM and the FNN, respectively, depending on the training size, i.e., the number of tracks. To enhance the statistical significance, for each data size reduction, the performance was evaluated by iterating over the complete dataset. Specifically, for the evaluation with only 10~$\%$ of the data, the same evaluation was done 10 times, to cover the entire input data. Mean values with related standard deviations are then shown in Figure~\ref{fig:result_scarcity}.
\\
\section{DISCUSSION}
\label{sec:discussion}
The results demonstrate that FNNs are capable of modeling PEs with the limited amount of available data. Correspondingly, the results indicate that FNNs are not universally superior to PMs. The general capability of FNNs to model any systematic behavior is well-established, as they function as universal function approximators \cite{cybenko1989approximation}. As stated in the introduction, this study serves as a proof-of-principle and focuses on exploring the practicality and efficiency of a (Blind-) Pointing Error Model involving FNNs. 

Comparisons of FNNs with multiple input features to the performance of a simple PM cannot provide insights in terms of a competitive analysis, as the PM operates on less information. The PM performance serves as a basic performance benchmark. Generally, adding more generic terms to the PM (Equations (\ref{equ:stdPM_dAZ}) and (\ref{equ:stdPM_dEL})) would transform the approach itself from a theoretically-based model towards a universal function approximator (e.g., Fourier series or orthogonal polynomials), making it more akin to the approach of an FNN. However, here, FNNs were the focus, and are particularly appealing due to their convenience in handling higher-dimensional input/output and their ability to model complex behavior.
\\
\subsection*{PM vs. FNN based on FS~P}
In two aspects of the shown results, the performance of the PM and the FNN can be compared directly: the final performance of both methods when evaluated/trained on the complete dataset using only FS~P as input (Figure~\ref{fig:result_ska_F11}), and the performance development with a variable amount of data for evaluation/training (Figure~\ref{fig:result_scarcity}), also based on FS~P.
The latter comparison reveals that the PM can be particularly superior to the FNN when less data is available for evaluation/training. Although the performances of the PM and FNN are quite similar when all 409 tracks are used for evaluation/training, the performance significantly diverges with a reduction in the underlying reference data size. For example, reducing the number of tracks to 204 (50~\%) decreases the performance of the PM by about 0.4~\%, for the FNN, it is reduced by about 8.3~\%. The discrepancy increases drastically with even smaller reference data sizes. While these results were qualitatively anticipated, it is beneficial to demonstrate the development in performance in this use case quantitatively. 

Figure~\ref{fig:result_scarcity} provides an empirical double-exponential fit for each model. Thus, stated values of convergence (PM: 15.6 arcsec, FNN: 16.7~arcsec) inherently possess a certain degree of uncertainty and should only serve as a guide to the eye. However, it can be concluded that the available data size of 409 tracks is just sufficient for the FNN to approximately match the PM's performance, with a potential decreasing trend for the FNN if more data (tracks) were used for training.
\\ 
\subsection*{Feature Selection with MI \& PC}
Selecting an effective and efficient set of features prior to training an FNN can significantly reduce computational time. Consequently, the idea naturally emerged to evaluate dependencies or correlations between potential input features and the PEs. MI was chosen due to its ability to indicate non-linear dependencies, while the PCC was chosen as a standard method for evaluating linear correlations. However, adding the features ranked according to their scores from the respective evaluation method (see Table~\ref{tab:ranked_MI_PCC}) cumulatively to the FS input of an FNN only partially showed the expected improvement in the FNN's performance (see Figure~\ref{fig:result_all_FS}~/ Table~\ref{tab:all_mean_results}). Overall, the selected features did improve the performance of the FNNs, but the ranking based on MI or PCC scores did not proportionally reflect their effectiveness in improving the FNN's performance. Interestingly, the highest-ranked features according to MI (v$_\text{AZ}$) and PCC (v$_\text{EL}$) analysis did not contribute to any improvements in performance. Nonetheless, the fact that adding the top four ranked features (MI~5~/ PC~5) led to an improvement of the results by 39~\% (MI) and 44~\% (PC), compared to the training on the basic input FS~P, confirms some effectiveness of the method. It is important to note that temperature, unsurprisingly, appears to be the primary driver of this improvement.
\\
\subsection*{Merging PM and FNN}
Contrasting the approach of substituting a PM with an FNN to model the complete PE, it appears more advantageous to merge both methodologies. The PM offers significant advantages, primarily by providing sound and robust performance with even a low amount of data for evaluation, due to its inherent information derived from physically based PE formulas. Here, a straightforward approach was tested in a two-step modeling architecture: first, the PM was evaluated on the training data to predict PEs for the test data. Discrepancies in these predictions represent residual PEs of the PM. Second, the FNN was trained on those residual PEs with various FSes as additional input. This ensures that the FNN does not expend parameters on replicating the PM's principles, but rather operates on more detailed or complex structures in the data.
The results in Table~\ref{tab:all_mean_results} are labelled \textit{PM~+ FNN} and demonstrate a consistent improvement in joint performance of 1.0~- 2.2 arcsec.

Out-of-scope outlook: to push the performance even further, the implemented PM should be extended, e.g., further terms for more variables in the input should be added (e.g., for the best-ranked PCC features) before training the FNN on its residuals. Preliminary tests indicate that this approach can achieve maximum performances of approximately 6.0~$\pm$~2.9~arcsec (with FS~X). 
\\
\section{CONCLUSION}
\label{sec:conclusion}
The results of this proof-of-principle study demonstrate that Feedforward Neural Networks (FNNs), even within the scope of the presented use case, can effectively function either as a substitute or a supplement to a Pointing Model (PM) to optimize the overall model's performance. However, numerous optimization opportunities remain unexplored. For example, most hyperparameters of the FNNs were not adjustable for optimizations, leaving potential improvements in both computational time and the final performance of the FNN untapped.
Additionally, the total number of parameters (weights and biases) required by the FNN has not been investigated, yet. On average, the best-performing FNNs in Table~\ref{tab:all_mean_results} utilized 657 parameters, whereas the median number of parameters was 347. Most certainly, there is potential for optimization in the architecture of FNN also for both the performance and required computational time \cite{eldan2016power}. 
Moreover, the area of feature selection and feature engineering presents an open field for improvements.
To name one example of feature engineering: calculating and adding the angle of attack from wind direction and the azimuth position of the telescope could be beneficial to evaluate quasi-static wind-induced Pointing Errors (PE) more efficiently. In terms of feature selection, it would be interesting to compare retrospective analysis methods on the importance of each feature, e.g., with SHapley
Additive exPlanations (SHAP) \cite{shapley1953value, lundberg2017unified, lundberg2018consistent}, Deep Learning Important FeaTures (DeepLIFT) \cite{shrikumar2017learning} or Local Interpretable Model-Agnostic Explanations (LIME) \cite{aditya2022local}. 

It should be emphasized that classical PMs are inherently robust and reliable for modeling PEs, especially when only a relatively small amount of reference data is available for evaluation. In contrast, FNNs are not tailored for predictions based on sparse data but are well-suited for modeling detailed and complex multidimensional data in a highly convenient manner. Once the software infrastructure is established, specifying input and output features is all that is required and can be applied to any use case. 

Aside from the technical aspects, considering that the available data was collected over a period of three weeks, it would be valuable to repeat this study with a larger dataset that comprises seasonal variations throughout the entire year.
\\
\acknowledgments 
This study has received funding from the European Union’s Horizon 2020 research and innovation program under grant agreement No. 951815 (AtLAST).\footnotemark[3]
\footnotetext[3]{\href{https://cordis.europa.eu/project/id/951815}{https://cordis.europa.eu/project/id/951815}}The consortium of AtLAST consists of the University of Oslo, the European Southern Observatory, OHB Digital Connect GmbH (formerly MT Mechatronics), the United Kingdom Astronomy Technology Centre (UK ATC), and the University of Hertfordshire. The authors wish to express their gratitude to AtLAST for the support and inspiration to enable this study. Additionally, we extend our appreciation to the Max Planck Institute for Radio Astronomy (MPIfR), particularly Dr. G. Wieching, for not only collecting the essential data for this study but also for the support in data handling by T. Glaubach prior to the analysis detailed in this manuscript. We also thank our colleague, A. Ippa, for the insightful discussions that significantly enriched this work.
\\
\bibliography{report} 
\bibliographystyle{spiebib} 

\end{document}